\documentclass[aps,prl,numerical,superscriptaddress,showpacs,floatfix,nofootinbib,reprint]{revtex4-1}
\usepackage{amsmath,amssymb,mathrsfs,bm}
\usepackage{url}
\usepackage{hyperref}
\hypersetup{
colorlinks=true,
linkcolor=blue,
anchorcolor =red,
citecolor=blue,
filecolor = red,
urlcolor=blue,
pdfauthor=author}
\usepackage{txfonts}
\usepackage[utf8]{inputenc}
\usepackage{amsmath}
\usepackage{graphicx}
\def\av<#1>{\left\langle\,#1\,\right\rangle}
\def\ev<#1>{\left\langle\,#1\,\right\rangle_{\rm{ev}}}

\begin{document}

%\pagewiselinenumbers

%%-----------------------------------------------------------
%\title{Do Small Systems Flow Like the QGP? \\ Species-Dependent Scaling of Anisotropic Flow Across Collision Systems}
%\title{Do QGP droplets develop in p+Au, d+Au and $^3$He+Au Collisions? \\ Species-Dependent Scaling of Anisotropic Flow Across Collision Systems }
%\title{Do Quark-Gluon Plasma Droplets Drive Azimuthal Anisotropy \\ in Small Systems at RHIC and the LHC?}
\title{Do QGP Droplets Drive Anisotropy in Small Systems? Insights from RHIC and the LHC}

\author{ Roy~A.~Lacey}
\email[E-mail: ]{Roy.Lacey@Stonybrook.edu}
\affiliation{Department of Chemistry, 
Stony Brook University, \\
Stony Brook, NY, 11794-3400, USA}
%
%\author{ N. Magdy}
%\email[E-mail: ]{Roy.Lacey@Stonybrook.edu}
%\affiliation{Department of Chemistry, 
%Stony Brook University, \\
%Stony Brook, NY, 11794-3400, USA}
%

\date{\today}

\begin{abstract}
Azimuthal anisotropy scaling functions for identified mesons and baryons are analyzed in large (Pb+Pb at \(\sqrt{s_{NN}} = 2.76\) and 5.02~TeV, Au+Au at \(\sqrt{s_{NN}} = 200\)~GeV), intermediate (Cu+Cu at \(\sqrt{s_{NN}} = 200\)~GeV), and small (p+Pb at \(\sqrt{s_{NN}} = 5.02\) and 8.16~TeV, p+Au, d+Au, and \(^3\)He+Au at \(\sqrt{s_{NN}} = 200\)~GeV) collision systems. The scaling functions’ fidelity supports a hydrodynamic-like origin for anisotropies in the flow-dominated regime. Central Pb+Pb, Au+Au, and Cu+Cu reflect QGP-driven expansion with strong radial flow and significant jet quenching, while peripheral Pb+Pb and Cu+Cu exhibit hadronic-dominated dynamics. In contrast, central RHIC small systems show hadronic-dominated behavior, with strong re-scattering, negligible radial flow, and suppressed jet quenching, following the hierarchy p+Au \(>\) d+Au \(>\) \(^3\)He+Au. At the LHC, ultra-central p+Pb collisions display enhanced radial flow, reduced re-scattering, and small but nonzero jet quenching. Scaling violations at high \(p_T\) reflect partial suppression of partonic energy loss. These findings demonstrate that QGP-like behavior in small systems depends sensitively on both system size and beam energy, and establish the scaling framework as a robust diagnostic of collectivity and medium properties across diverse collision conditions.
\end{abstract}

%\begin{keyword}
%Quark--gluon plasma \sep transport coefficients \sep Ultra-central heavy-ion collisions
%%% PACS codes here, in the form: \PACS code \sep code
%\end{keyword}

\maketitle

%%-----------------------------------------------------------
%
%\section{Introduction}
Understanding the emergent properties of strongly interacting matter under extreme 
conditions is a central goal of quantum chromodynamics (QCD)~\cite{Shuryak:1980tp}. 
High-energy heavy-ion collisions provide a unique means to create and study the 
quark-gluon plasma (QGP)—a deconfined phase of quarks and gluons believed to have 
existed microseconds after the Big Bang~\cite{Yagi:2005yb,Gyulassy:2004zy}. 
Among proposed QGP signatures, azimuthal anisotropy has proven especially sensitive 
to collective behavior and transport properties~\cite{Heinz:2013th}.

Azimuthal anisotropy is quantified as
\begin{equation}
    \frac{dN}{d\phi} \propto 1 + \sum_{n=1}^{\infty} 2v_n(p_T) \cos[n(\phi - \Psi_n)],
\end{equation}
where \(\phi\) is the azimuthal angle, \(\Psi_n\) the event-plane angle, 
and \(v_n\) the anisotropy coefficient. These coefficients probe both transport 
properties—such as the specific shear viscosity \((\eta/s)\)—and thermodynamic 
features of the equation of state (EOS)~\cite{Ollitrault:1992bk,Voloshin:1994mz,
Danielewicz:1998vz,Voloshin:2008dg,Heinz:2013th,Gale:2013da}. 
At low transverse momentum \(p_T\), \(v_n(p_T, \text{cent})\) reflects the hydrodynamic 
response to initial-state geometry~\cite{Gale:2013da,Heinz:2013th}. 
At high \(p_T\), it is dominated by path-length-dependent parton energy loss, 
governed by the jet-quenching parameter \(\hat{q}\)~\cite{Baier:1996kr,Majumder:2011uk,
Salgado:2003qc,JETSCAPE:2020mzn}, thereby constraining the medium’s density and opacity.

In large collision systems (e.g., Au+Au at \(\sqrt{s_{NN}} = 200\)~GeV, 
Pb+Pb at \(\sqrt{s_{NN}} = 5.02\)~TeV), the long-lived QGP exhibits both 
strong collectivity and significant jet quenching, enabling detailed studies 
of its transport and thermodynamic properties.

Significant azimuthal anisotropy has also been observed in small-system collisions, such as p+Au, d+Au, and \(^{3}\)He+Au at \(\sqrt{s_{NN}} = 200\)~GeV, and p+Pb at \(\sqrt{s_{NN}} = 5.02\) and 8.16~TeV~\cite{PHENIX:2017djs,STAR:2023wmd,CMS:2012qk,CMS:2018loe,ALICE:2012eyl,ATLAS:2012cix}. These results challenge expectations for the conditions required to produce QGP-like behavior. Low-\(p_T\) anisotropies are often interpreted as collective flow driven by initial geometry, but in small systems it remains unclear whether they arise from QGP-driven hydrodynamics or from hadronic or non-QGP effects. This ambiguity raises two central questions:
\begin{enumerate}
    \item Do the low-\(p_T\) anisotropies in small systems reflect genuine hydrodynamic expansion, or could they stem from other mechanisms such as initial-state correlations (e.g., gluon saturation in the color-glass condensate) or parton escape dynamics~\cite{Dusling:2015gta,Schlichting:2016sqo}?
    \item If hydrodynamic behavior is present, do the short lifetimes, low multiplicities, and large surface-to-volume ratios limit QGP development? Would this favor hadronic re-scattering and suppress jet quenching at high \(p_T\), altering the balance between flow- and jet-driven anisotropies and complicating their interpretation?
\end{enumerate}

Azimuthal anisotropy scaling functions for identified particles provide a unified framework to examine the system-size and energy dependence of \(v_2(p_T, \text{cent})\) and \(v_3(p_T, \text{cent})\)~\cite{Lacey:2013qua,Lacey:2024uky,Lacey:2024bcm}. These functions incorporate initial-state eccentricities (\(\varepsilon_n\)), specific shear viscosity (\(\eta/s\)), stopping power (\(\hat{q}\)), the dimensionless system size (\(\mathbb{R} \propto RT\)), radial flow, and viscous corrections to the distribution function (\(\delta_f\))~\cite{Majumder:2007zh,Dusling:2009df}. Here, \(R\) denotes the transverse size and \(T\) the freeze-out temperature, characterizing geometric and thermal properties. By linking \(\eta/s\) to both flow and energy-loss dynamics, scaling functions yield predictive patterns that distinguish hydrodynamic behavior from alternatives such as initial-state correlations or parton escape~\cite{Lacey:2013qua,Lacey:2024uky,Lacey:2024bcm}. For hydrodynamic-like evolution, they disentangle radial flow, viscous attenuation, hadronic re-scattering, and jet quenching, providing a data-driven probe of their interplay across systems and energies.

In large systems at high energy, a strongly interacting QGP accounts for the observed anisotropies. At low \(p_T\), collective expansion dominates with minimal hadronic re-scattering~\cite{Teaney:2001av,Huovinen:2001cy,Gale:2013da,Heinz:2013th,Lacey:2024uky}, while strong \(v_n\) signals and robust scaling~\cite{Lacey:2013qua,Lacey:2024uky,Lacey:2024bcm} support this hydrodynamic picture. The coexistence of strong radial flow at low \(p_T\) and jet quenching at high \(p_T\) is a hallmark of QGP: a medium both low in viscosity and highly opaque to energetic partons.

At lower energies, the dynamics shift toward a hadron-dominated regime with increased re-scattering, longer hadronic lifetimes, and a modest (15–20\%) rise in effective \(\eta/s\) as \(\sqrt{s_{NN}}\) decreases from $\sim 39$ to 11.5~GeV~\cite{Lacey:2024bcm}. In small systems, short lifetimes, low multiplicities, and large surface-to-volume ratios suggest a reduced or absent QGP phase. The interplay among radial flow, viscous attenuation, re-scattering, and quenching remains unresolved, yet its resolution can provide detailed insight into the QGP’s role. Scaling functions offer a direct, data-driven means to disentangle these effects—especially in the transition from low-\(p_T\) (flow-driven) to high-\(p_T\) (quenching-driven) regimes—clarifying how collectivity and medium properties evolve with system size and collision energy.

This study presents $v_2(p_T, \text{cent})$ scaling functions for charged hadrons ($h^{\pm}$), neutral kaons ($K^0$), and lambda baryons ($\Lambda\bar{\Lambda}$) in p+Pb collisions at $\sqrt{s_{NN}} = 5.02$~TeV and Pb+Pb collisions at $\sqrt{s_{NN}} = 2.76$~TeV, matched by multiplicity. Additional analyses include identified pions, kaons, and protons in p+Au, d+Au, $^{3}$He+Au, Cu+Cu, and Au+Au at $\sqrt{s_{NN}} = 200$~GeV, as well as Pb+Pb at $\sqrt{s_{NN}} = 5.02$~TeV. Scaling comparisons at $\sqrt{s_{NN}} = 8.16$~TeV in p+Pb further incorporate heavier species, including $D^0$, $\Xi^{\pm}$, and $\Omega^{\pm}$.

Together, these measurements provide a comprehensive dataset for exploring azimuthal anisotropy across systems of varying size, geometry, and energy. The scaling framework enables systematic comparisons of contributions from shear viscosity, radial flow, hadronic re-scattering, and jet quenching, distinguishing QGP-driven anisotropies in large systems from hadronic-dominated behavior in smaller systems, and offering quantitative insight into the mechanisms of collectivity across a broad range of collision conditions.

Species-resolved scaling functions follow established procedures for charged hadrons and identified particles~\cite{Lacey:2024fpb,Lacey:2024uky}. Mesons and baryons share a common attenuation parameter \(\beta\), which encodes viscous damping and scales with the specific shear viscosity \((\eta/s)\). Since \(\eta/s \propto T^3/\hat{q}\), \(\beta\) is also sensitive to the jet-quenching coefficient \(\hat{q}\). Additional inputs are the initial eccentricities \(\varepsilon_n\); a transverse-size proxy \(\mathbb{R} \propto \langle N_{\text{chg}} \rangle_{|\eta|\le 0.5}^{1/3}\); and two species- and system-dependent parameters: \(\zeta_{\rm rf}^{(X)}\) for radial-flow blue shift and \(\zeta_{\rm hs}^{(X)}\) for hadronic re-scattering. The viscous correction is implemented as \(\delta_f = \kappa p_T^2\) with \(\kappa = 0.17~(\text{GeV}/c)^{-2}\)~\cite{Majumder:2007zh,Dusling:2009df,Liu:2018hjh}.

Centrality-dependent charged-particle multiplicities \(\langle N_{\text{chg}} \rangle_{|\eta|\le 0.5}\) are taken from published data~\cite{ALICE:2015juo,Lacey:2016hqy,ALICE:2012eyl,CMS:2013jlh,PHENIX:2021dod}. All scaling relations are defined relative to a common baseline: charged kaons in ultra-central Pb+Pb at \(\sqrt{s_{NN}} = 5.02\) TeV. Kaons serve as a minimally re-scattered reference due to their intermediate mass and small hadronic cross section. The baseline attenuation is \(\beta_0\), with system and energy dependence encoded as \(\beta = k_\beta \cdot \beta_0\).

To compare hadrons of different mass, transverse kinetic energy \({\rm KE}_T = m_T - m_0\) (with \(m_T = \sqrt{p_T^2 + m_0^2}\)) is used as the abscissa. This choice minimizes trivial kinematic effects, highlights collective flow, and ensures consistent treatment of mesons and baryons.

The meson \(v_2\) scaling relation is
\begin{multline}
\frac{v_2(p_T, \text{uc})}{\varepsilon_2(\text{uc})}
\, e^{ \tfrac{2 \beta_0}{\mathbb{R}_{\rm uc}} (2 + \kappa p_T^2) } \\
= \;
e^{ \alpha \,\zeta^{(X)}_M \tfrac{2 \beta_0}{\mathbb{R}_{\rm uc}} (2 + \kappa p_T^2) } \;
\left( \frac{v_2'(p_T)}{\varepsilon_2'} \right)^{\zeta^{(X)}_m} \;
e^{ \tfrac{2 \zeta^{(X)}_m \beta}{\mathbb{R}_{\rm uc}} 
\left( \tfrac{\mathbb{R}_{\rm uc}}{\mathbb{R}'} - 1 \right)(2 + \kappa p_T^2) } ,
\label{eq:v2_scaling_mesons}
\end{multline}
where primed quantities denote the comparison system and centrality. The attenuation factor is \(\zeta^{(X)}_m = 1 - \zeta_{\rm hs}^{(X)}\), with \(\zeta_{\rm hs}^{(X)}\) encoding hadronic re-scattering. For heavier mesons such as the \(\phi\), where \(\zeta_{\rm hs}^{(\phi)} \approx 0\), residual deviations from kaon scaling mainly reflect mass-dependent radial flow.

The normalization exponent is \(\zeta^{(X)}_M = (\zeta^{(X)}_m + \gamma_{32}) + (1 - k_\beta)\), with \(\gamma_{32} = 0\) for the kaon reference. The coefficient \(\alpha=1\) for ultra-central, \(\alpha \approx 0.56\) for near-ultra-central (0–5\%), and is absorbed for non-ultra-central collisions, where the \((2 + \kappa p_T^2)\) prefactor is omitted. Geometric factors \(\varepsilon_n\) and \(\mathbb{R} \propto \langle N_{\text{chg}} \rangle^{1/3}_{|\eta|\le 0.5}\) enter consistently, with \(\mathbb{R}_{\rm uc}\) the ultra-central reference. The prefactor \((2 + \kappa p_T^2)\) captures both inviscid and viscous response.

The baryon \(v_2\) scaling relation mirrors the meson case:
\begin{multline}
\frac{v_2(p_T, \text{uc})}{\varepsilon_2(\text{uc})}
\, e^{ \tfrac{2 \beta_0}{\mathbb{R}_{\rm uc}} (2 + \kappa p_T^2) } \\
= \;
e^{ \alpha \,\zeta^{(X)}_B \tfrac{2 \beta_0}{\mathbb{R}_{\rm uc}} (2 + \kappa p_T^2) } \;
\left( \frac{v_2'(p_T)}{\varepsilon_2'} \right)^{\zeta_b^{(X)}} \;
e^{ \tfrac{2 \zeta_b^{(X)} \beta}{\mathbb{R}_{\rm uc}}
\left( \tfrac{\mathbb{R}_{\rm uc}}{\mathbb{R}'} - 1 \right) (2 + \kappa p_T^2) } ,
\label{eq:v2_scaling_baryons}
\end{multline}
with \(\zeta_b^{(X)} = (1 - \zeta_{\rm rf}^{(X)})^{|n_B^X|}\), where \(\zeta_{\rm rf}^{(X)}\) encodes radial-flow blue shift and \(n_B^X\) is the baryon number. The normalization exponent is
\(\zeta^{(X)}_B = \zeta_b^{(X)} - \left(1/k_{\beta} - \gamma_{32}\right) - (1 - k_\beta)\), with \(\zeta_B = 0\) for the reference system.

The coefficient \(\alpha\) follows the meson convention: unity for ultra-central, \(\approx 0.56\) for near-uc, and absorbed for non-uc collisions, where the \((2 + \kappa p_T^2)\) prefactor is omitted. The viscous correction \(\delta_f = \kappa p_T^2\) applies uniformly to mesons and baryons, governing attenuation over the full \(p_T\) range and fixed above \(p_T^{\rm thresh} \sim 4.5\)~GeV/\(c\), where partonic energy loss dominates.

Together, Eqs.~\eqref{eq:v2_scaling_mesons}--\eqref{eq:v2_scaling_baryons} define a unified scaling framework that constrains \(\eta/s\) and \(\hat{q}\) across systems, centralities, and energies, while leveraging species-dependent differences in \(v_n\) to quantify radial flow and hadronic re-scattering.

\begin{figure*}[htb]
  \centering
    \includegraphics[width=0.75\textwidth]{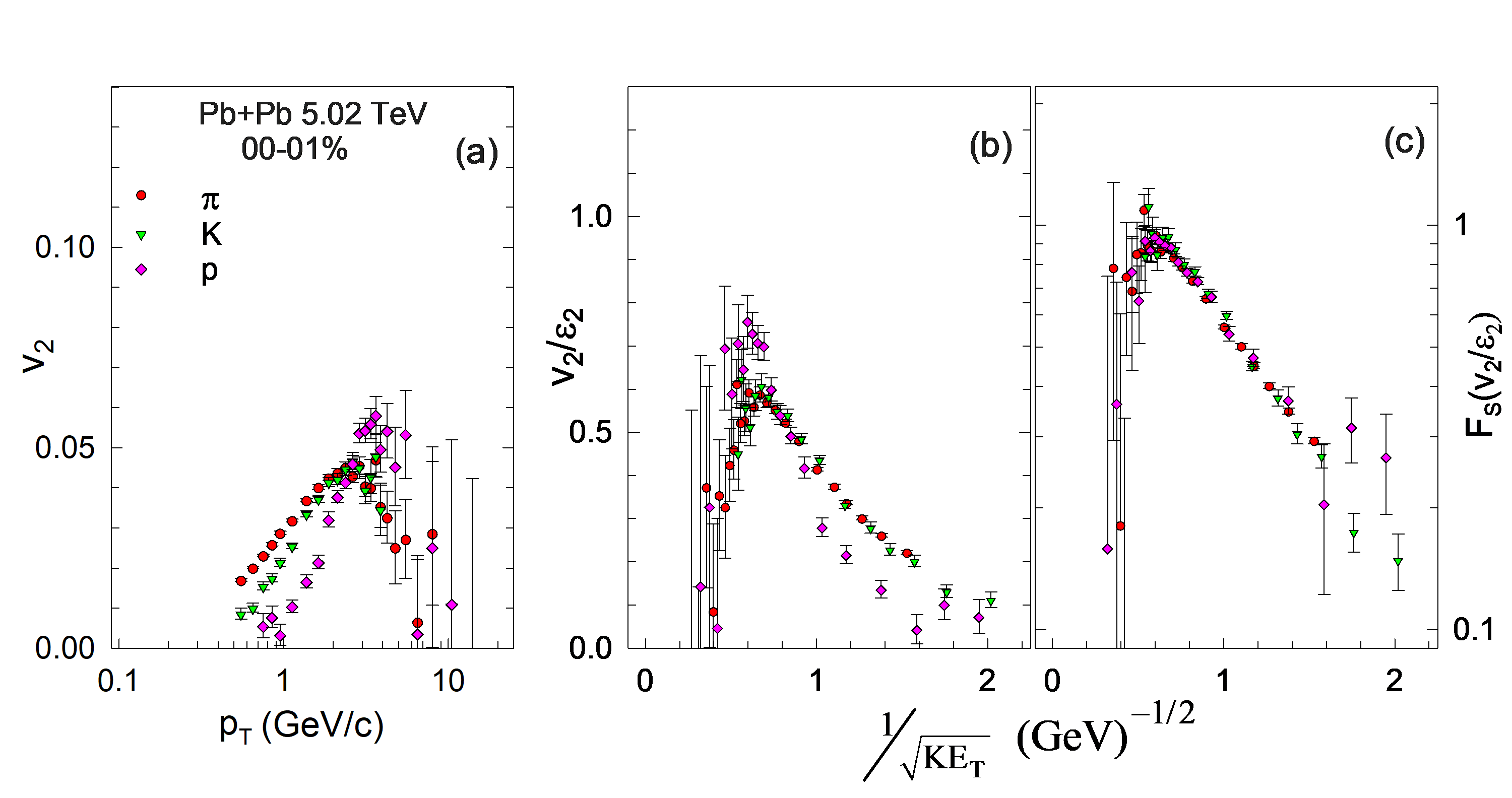} \\
    \vspace{-0.18cm}
    \includegraphics[width=0.75\textwidth]{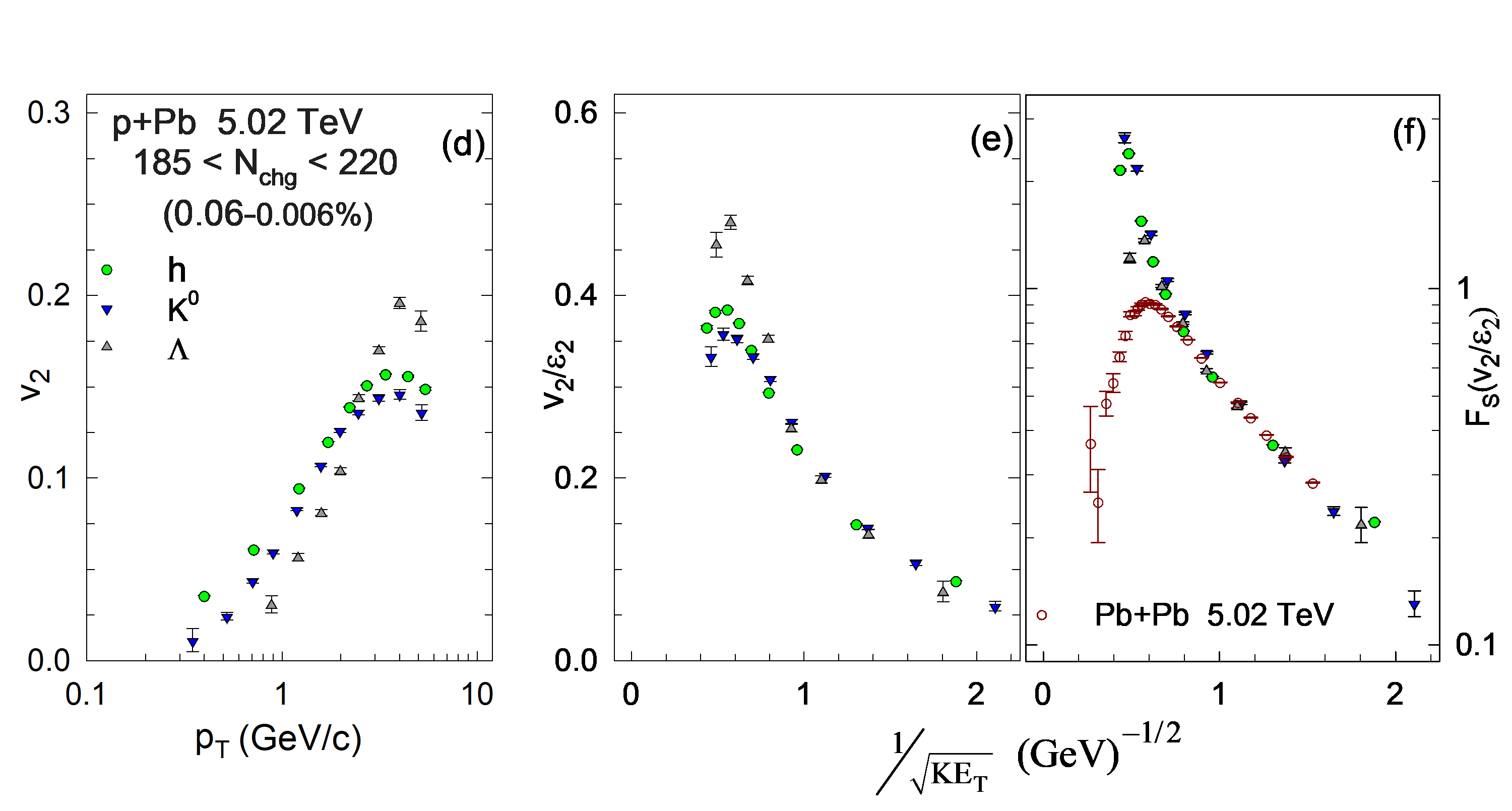} \\
    \vspace{-0.2cm}
    \caption{(Color online) Construction of anisotropy scaling functions for identified species in ultra-central Pb+Pb and p+Pb collisions at \(\sqrt{s_{NN}} = 5.02\)~TeV. 
Top panels: \(\pi^{\pm}\), \(K^{\pm}\), and \(p\bar{p}\) in Pb+Pb. Bottom panels: \(h^{\pm}\), \(K^0\), and \(\Lambda\bar{\Lambda}\) in p+Pb, with inclusive hadrons as pion proxies. 
Panels (a,d) show raw \(v_2(p_T)\), (b,e) the eccentricity-scaled values \(v_2/\varepsilon_2\), and (c,f) the final scaling functions vs.\ \(1/\sqrt{{\rm KE}_T}\). 
Species convergence in the flow-dominated region demonstrates common medium response, while deviations at high \(p_T\) indicate reduced opacity and suppressed jet quenching in smaller systems. 
Data from ALICE and CMS~\cite{ALICE:2022zks,CMS:2014und,CMS:2018loe}.}
\label{fig1}
\end{figure*}
\begin{figure*}[htb]
    \centering
    \includegraphics[width=0.75\textwidth]{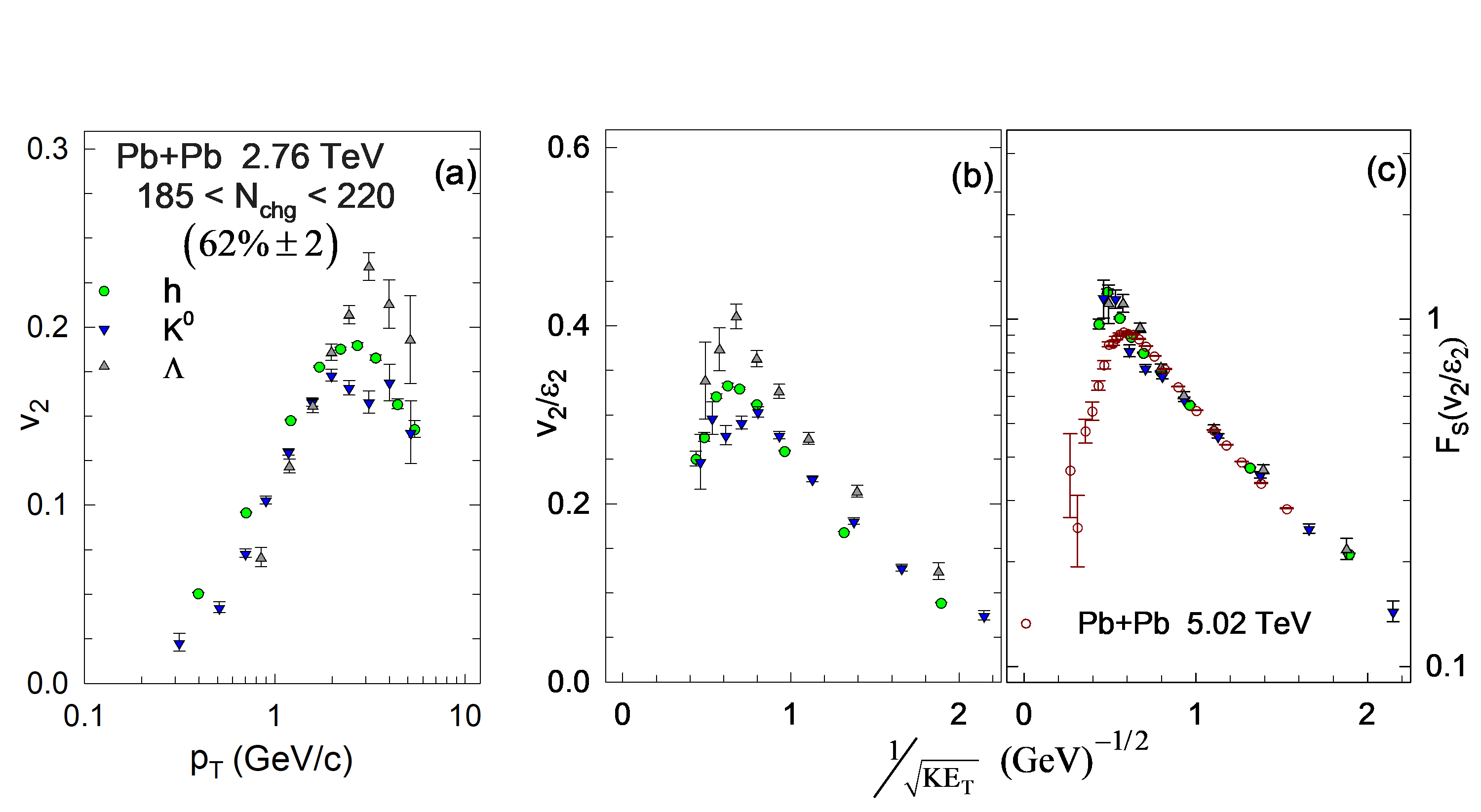} \\
    \vspace{-0.18cm}
    \includegraphics[width=0.75\textwidth]{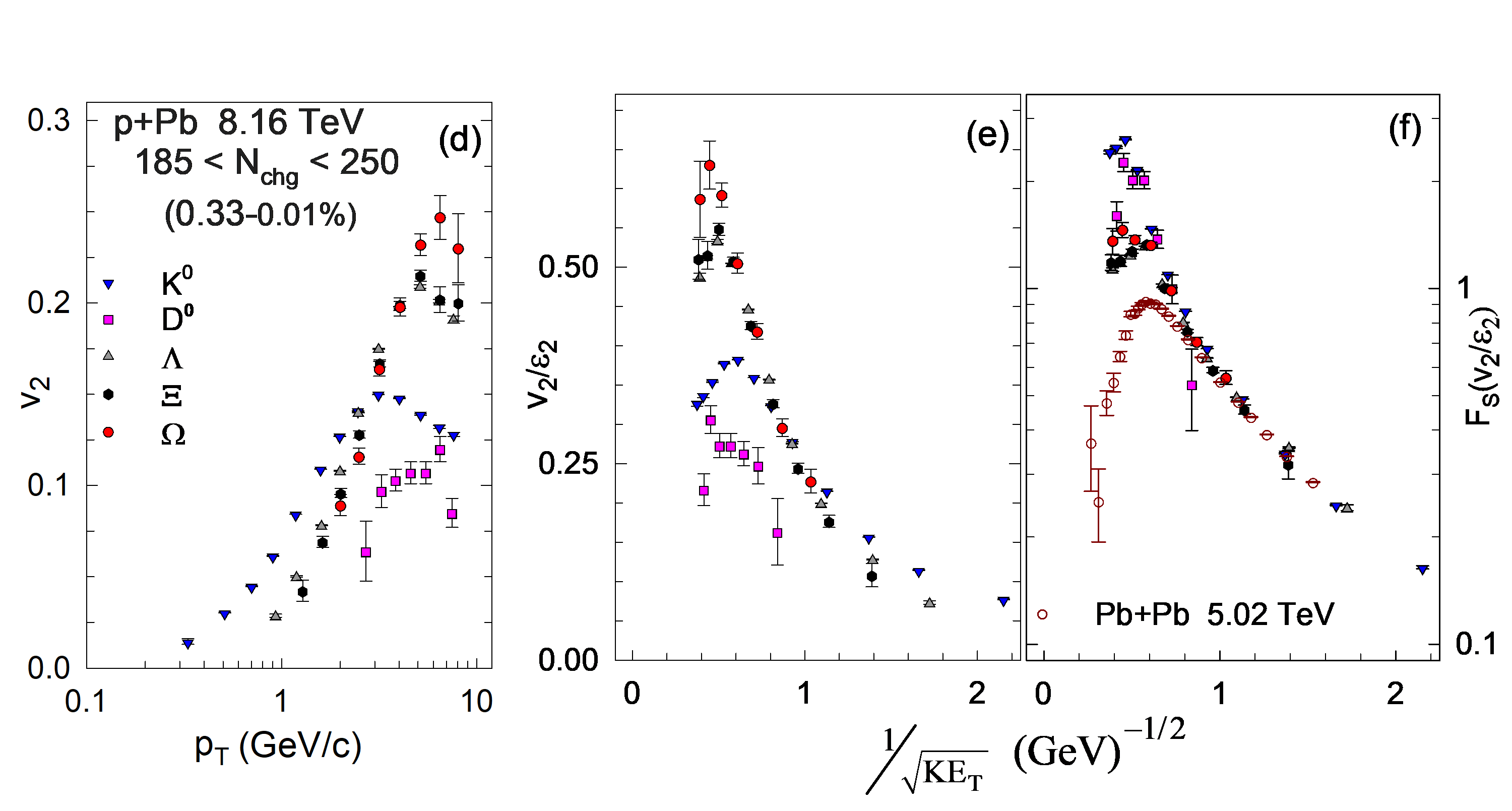}
    \vspace{-0.4cm}
    \caption{(Color online) Comparison of scaling functions for identified particles in peripheral Pb+Pb (\(\sqrt{s_{NN}} = 2.76\)~TeV) and ultra-central p+Pb (\(\sqrt{s_{NN}} = 8.16\)~TeV) at matched multiplicity. 
Top panels: \(h^{\pm}\), \(K^0\), and \(\Lambda\bar{\Lambda}\) in Pb+Pb (inclusive hadrons as pion proxies). 
Bottom panels: p+Pb, including \(D^0\), \(\Xi^{\pm}\), and \(\Omega^{\pm}\). 
Panels (a,d) show raw \(v_2(p_T)\); (b,e) the eccentricity-scaled values \(v_2/\varepsilon_2\); and (c,f) the final scaling functions. 
Species convergence in the flow-dominated region indicates common medium response, while high-\(p_T\) deviations signal reduced opacity and suppressed jet quenching in small systems. 
Data from ALICE and CMS~\cite{ALICE:2022zks,CMS:2014und,CMS:2018loe}.}
    \label{fig2}
\end{figure*}

Initial-state eccentricities \(\varepsilon_n\) were calculated with the Monte Carlo quark-Glauber (MC-qGlauber) model~\cite{Liu:2018hjh}, an extension of the standard MC-Glauber~\cite{Miller:2007ri,PHOBOS:2006dbo} that includes spatial quark distributions within nucleons and inelastic cross sections consistent with nucleon--nucleon data. This refinement accounts for nucleon size, internal structure, wounding profiles, and system-dependent nuclear configurations. Eccentricity fluctuations are implemented for all systems, with simulations tailored to Au and Pb nuclei for large systems and to asymmetric projectiles (p, d, \(^{3}\)He) for small systems. These fluctuations play a central role in small systems, where the observed anisotropy is driven largely by event-by-event eccentricity fluctuations rather than average geometry. The systematic uncertainty on \(\varepsilon_n\) is about 2\%, based on model variations.

The \(v_2(p_T, \text{cent})\) data used to construct the scaling functions include \(h^{\pm}\), \(\pi^{\pm}\), \(K^0\), \(K^{\pm}\), \(D^0\), \(p\bar{p}\), \(\Lambda\bar{\Lambda}\), \(\Xi^{\pm}\), and \(\Omega^{\pm}\), reported by ALICE, CMS, and PHENIX~\cite{ALICE:2022zks,CMS:2014und,CMS:2018loe,PHENIX:2014uik,PHENIX:2014yml,PHENIX:2017djs}. Particle-plus-antiparticle averages are used where appropriate. By scaling results from small systems (p+Pb, p+Au, d+Au, \(^{3}\)He+Au), large systems (Pb+Pb, Au+Au), and the intermediate Cu+Cu system at \(\sqrt{s_{NN}} = 200\)~GeV, the analysis enables direct comparison of viscous attenuation, radial flow, hadronic re-scattering, and partonic energy loss across diverse system sizes and energies (200~GeV, 2.76~TeV, 5.02~TeV, and 8.16~TeV). For Cu+Cu, peripheral centrality selections yield multiplicities comparable to those in \(^3\)He+Au, providing a natural bridge between asymmetric small systems and larger nuclei—analogous to matched-multiplicity comparisons of p+Pb and Pb+Pb at the LHC.

Figure~\ref{fig1} illustrates the scaling procedure for ultra-central Pb+Pb and p+Pb at \(\sqrt{s_{NN}} = 5.02\)~TeV. Panels (a) and (d) show raw \(v_2(p_T)\): \(\pi^{\pm}\), \(K^{\pm}\), and \(p\bar{p}\) for Pb+Pb (a), and \(h^{\pm}\), \(K^0\), and \(\Lambda\bar{\Lambda}\) for p+Pb (d), with inclusive hadrons used as pion proxies~\cite{CMS:2013pdl,ALICE:2013wgn}. Both systems show the expected species ordering, but \(v_2\) is larger in p+Pb, consistent with its greater initial eccentricity. To separate flow- and quenching-dominated regions, panels (b) and (e) plot \(v_2/\varepsilon_2\) versus \(1/\sqrt{{\rm KE}_T}\), while panels (c) and (f) show the fully scaled results~\cite{Dokshitzer:2001zm,Lacey:2010fe}.

Eccentricity scaling with transverse kinetic energy improves meson agreement [panels (b,e)], emphasizing the geometric origin of anisotropy, but blue shifts remain between mesons and baryons—especially in Pb+Pb—signaling stronger radial flow. In the fully scaled Pb+Pb results [panel (c)], all species collapse onto a universal curve, confirming robust scaling. Extracted parameters, \(\alpha = 1\), \(\beta_0 = 0.88\), \(k_{\beta} = 1.0\) (\(\beta=\beta_0\)), \(\zeta_{\rm rf} = 0.49\), \(\zeta_{\rm hs} = 0.00\), and \(\gamma_{32}=0\), indicate QGP-dominated physics: low-viscosity hydrodynamics (\(\eta/s\)), strong partonic energy loss (\(\hat{q}\)), and large radial flow. The vanishing \(\zeta_{\rm hs}\) reflects negligible hadronic re-scattering. A full summary is given in Table~\ref{tab:coefficients1}.

In p+Pb [panel (f)], scaling collapse holds at low \(p_T\) but breaks at intermediate and high \(p_T\). These violations lessen as the analysis threshold is reduced from the canonical \(p_T^{\rm thresh}\!\sim\!4.5\)~GeV/\(c\), consistent with reduced opacity and short QGP lifetimes. Jet quenching appears strongly suppressed but not absent. If violations persist at high \(p_T\) even after threshold reduction, they would instead signal unquenched-jet contributions. Unlike conventional subtraction using \(p+p\) references—which can overestimate non-flow at low \(p_T\)—the present framework interprets such residuals directly as jet-related correlations.

Extracted parameters for p+Pb, \(k_{\beta} = 1.0\), \(\zeta_{\rm rf} = 0.26\), \(\zeta_{\rm hs} = 0.11\), and \(\gamma_{32}=0\), show no change in \(\eta/s\) relative to Pb+Pb, a modest rise in hadronic re-scattering (with uncertainties from inclusive-hadron use), and reduced radial flow linked to weaker opacity. These contrasts suggest that QGP effects remain in ultra-central p+Pb but are markedly weaker than in Pb+Pb collisions.

Figure~\ref{fig2} compares scaling for Pb+Pb at \(\sqrt{s_{NN}} = 2.76\)~TeV and p+Pb at \(\sqrt{s_{NN}} = 8.16\)~TeV, matched by multiplicity to probe systems with similar particle yields but different geometry, size, and beam energy. Panels (a–c) show \(h^{\pm}\), \(K^0\), and \(\Lambda\bar{\Lambda}\) in peripheral (62\%) Pb+Pb, using inclusive hadrons as pion proxies; panels (d–f) show \(K^0\), \(D^0\), \(\Xi^{\pm}\), and \(\Omega^{\pm}\) in ultra-central p+Pb. The ultra-central p+Pb data at 5.02~TeV from Fig.~\ref{fig1} also fall in this multiplicity class, enabling broader cross-system comparisons.

Panels (a,d) present raw \(v_2(p_T)\). Overall magnitudes are similar in Pb+Pb and p+Pb, implying comparable initial eccentricities despite the latter’s much smaller size. This may reflect enhanced eccentricity fluctuations in p+Pb, capable of generating large anisotropies even with little geometric eccentricity. Species separations are more pronounced in p+Pb: the wider \(K^0\)–\(\Lambda\) splitting and sizable \(\Xi\), \(\Omega\) separations point to stronger radial flow, likely from steeper pressure gradients in its compact, high-density geometry. For \(D^0\), the limited \(p_T\) coverage makes it difficult to discern the trend.

Panels (b,e) show eccentricity-scaled \(v_2\) vs.\ \(1/\sqrt{{\rm KE}_T}\). 
In peripheral Pb+Pb (panel b), scaling improves the agreement among meson species, 
while in p+Pb (panel e) it enhances the alignment among baryon species. 
The larger baryon–meson blue shift in p+Pb relative to peripheral Pb+Pb is consistent with 
stronger pressure gradients in the smaller system. 
These results highlight the importance of initial geometry and fluctuations and 
show that small systems can exhibit collective behavior consistent with hydrodynamic-like expansion.

Panels (c,f) display the fully scaled results. Species collapse in the flow-dominated domain demonstrates robust scaling across systems matched in multiplicity but differing in geometry and energy. Violations persist at intermediate and high \(p_T\), especially in p+Pb, but diminish as the threshold is lowered from \(p_T^{\rm thresh}\!\sim\!4.5\)~GeV/\(c\). This suggests jet quenching is suppressed but not absent, while residual high-\(p_T\) violations likely signal non-flow contributions from unquenched jets.

The observed suppression hierarchy—uc p+Pb (5.02–8.16~TeV) \(>\) Pb+Pb (62\%, 2.76~TeV) \(>\) uc Pb+Pb (5.02~TeV)—encodes the combined influence of system size, path length, and QGP lifetime. These trends constrain opacity-dependent jet quenching and highlight the value of matched-multiplicity comparisons for disentangling flow- and jet-driven effects.
\begin{table}[h!]
\centering
\caption{Extracted scaling coefficients for Pb+Pb and p+Pb collisions. 
The value \(\beta_0 = 0.88\) for Pb+Pb at \(5.02\)~TeV is used to compute 
\(k_{\beta} = \beta'/\beta_0\). Peripheral Pb+Pb and ultra-central p+Pb 
have comparable multiplicities.}
\label{tab:coefficients1}
\begin{tabular}{lccccc}
\hline
System & Centrality (\%) & \(\gamma_{32}\) & \(k_{\beta}\) & \(\zeta_{\rm hs}\) & \(\zeta_{\rm rf}\) \\
\hline
Pb+Pb (5.02 TeV) & \(0-1\)    & 0.0 & 1.00 & 0.00 & 0.48 \\
Pb+Pb (2.76 TeV) & \(0-1\)    & 0.0 & 0.95 & 0.00 & 0.47 \\
Pb+Pb (2.76 TeV) & \(62\pm 2\) & 0.0 & 1.00 & 0.14 & 0.02 \\
p+Pb (5.02 TeV)  & \(0.006-0.06\) & 0.0 & 1.00 & 0.11 & 0.26 \\
p+Pb (8.16 TeV)  & \(0.01-0.33\) & 0.0 & 1.00 & --   & 0.33 \\
\hline
\end{tabular}
\end{table}

Table~\ref{tab:coefficients1} lists the extracted scaling parameters: 
\(\zeta_{\rm hs}\) and \(\zeta_{\rm rf}\) quantify hadronic re-scattering and radial flow, 
while \(k_\beta = \beta'/\beta_0\) captures viscous attenuation. 
The results confirm strong QGP-driven dynamics in uc Pb+Pb, with substantial jet quenching, 
but show that peripheral Pb+Pb expansion is dominated by hadronic re-scattering rather than QGP evolution. 
Ultra-central p+Pb exhibits intermediate behavior—reduced radial flow, modest re-scattering, 
and suppressed jet quenching—consistent with short-lived but non-negligible QGP evolution. 
Notably, uc p+Pb develops stronger radial flow than peripheral Pb+Pb, 
likely from steep pressure gradients in its compact, high-density geometry.

\begin{figure*}
  \includegraphics[width=0.70\textwidth]{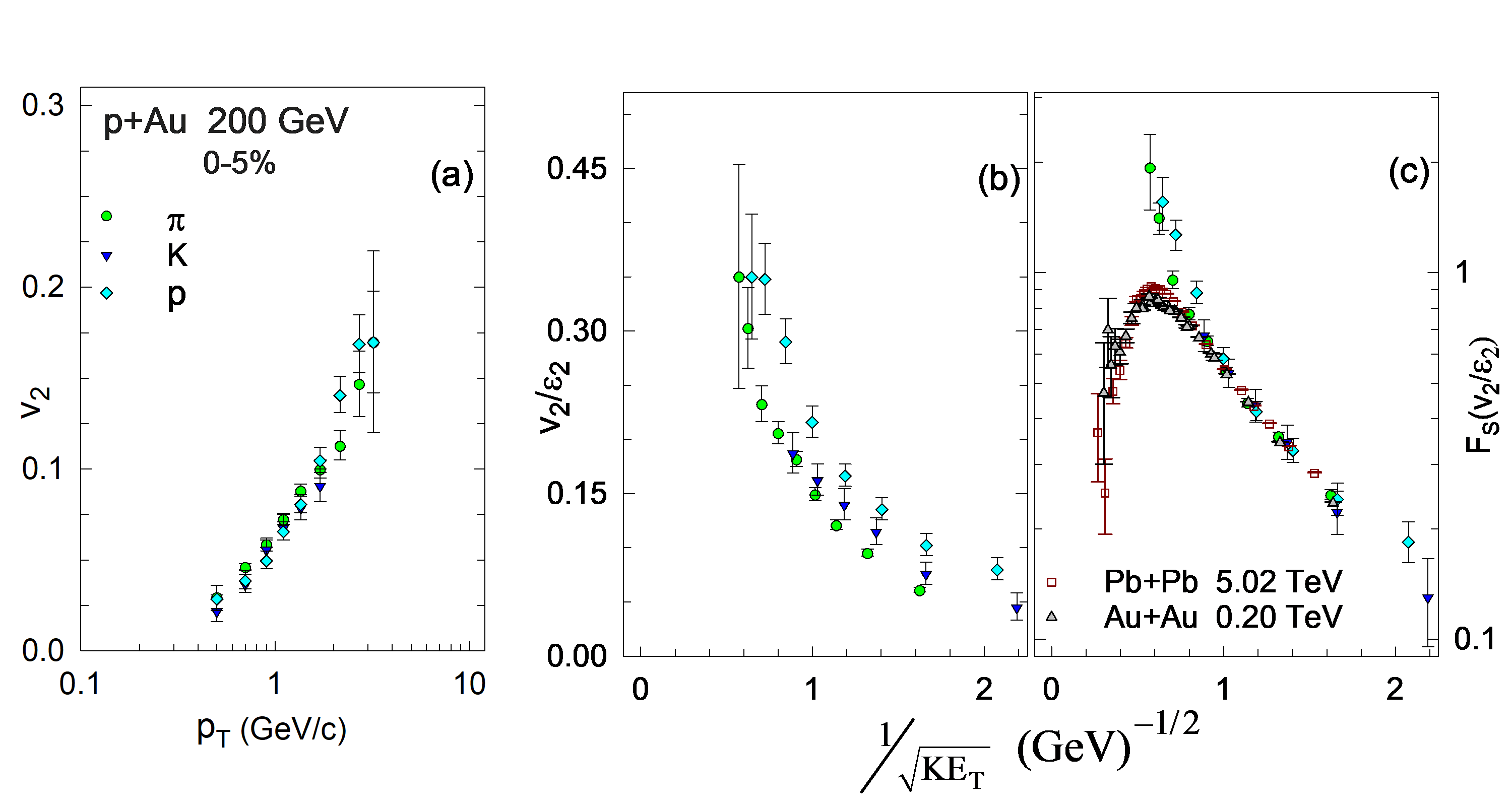}
	 \vskip -0.3 cm
  \caption{(Color online) Scaling procedure for 0–5\% central p+Au collisions at \(\sqrt{s_{NN}} = 200\)~GeV. 
Panel (a): raw \(v_2(p_T)\) for mesons (\(\pi^{\pm}, K^{\pm}\)) and baryons (\(p\bar{p}\)). 
Panel (b): eccentricity-scaled values \(v_2/\varepsilon_2\). 
Panel (c): final scaling function, compared with central Pb+Pb (\(\sqrt{s_{NN}} = 5.02\)~TeV) and Au+Au (\(\sqrt{s_{NN}} = 200\)~GeV). 
Data from ALICE and PHENIX~\cite{ALICE:2022zks,PHENIX:2017djs}.}
  \label{fig3}
\end{figure*}

%

%\begin{figure*}
  %\includegraphics[width=0.70\textwidth]{v2_pid_00-05-dAu0.02TeV-scaling.PNG}
	 %\vskip -0.3 cm
  %\caption{(Color Online) Comparison of \(v_2(p_T)\) for mesons (\(\pi^{\pm}\)) and baryons (\(p\bar{p}\)) in panel (a), their eccentricity-scaled values \(v_2/\varepsilon_2\) in panel (b), and the resulting scaling function in panel (c) for 0–5\% central d+Au collisions at \(\sqrt{s_{NN}} = 200\) GeV. The scaling function for central Pb+Pb collisions at \(\sqrt{s_{NN}} = 5.02\) TeV is also shown in panel (c) for comparison. The data are sourced from the ALICE and PHENIX collaborations \cite{ALICE:2022zks,PHENIX:2017djs}.}
  %\label{fig4}
%\end{figure*}
%

%
\begin{figure*}
  \includegraphics[width=0.70\textwidth]{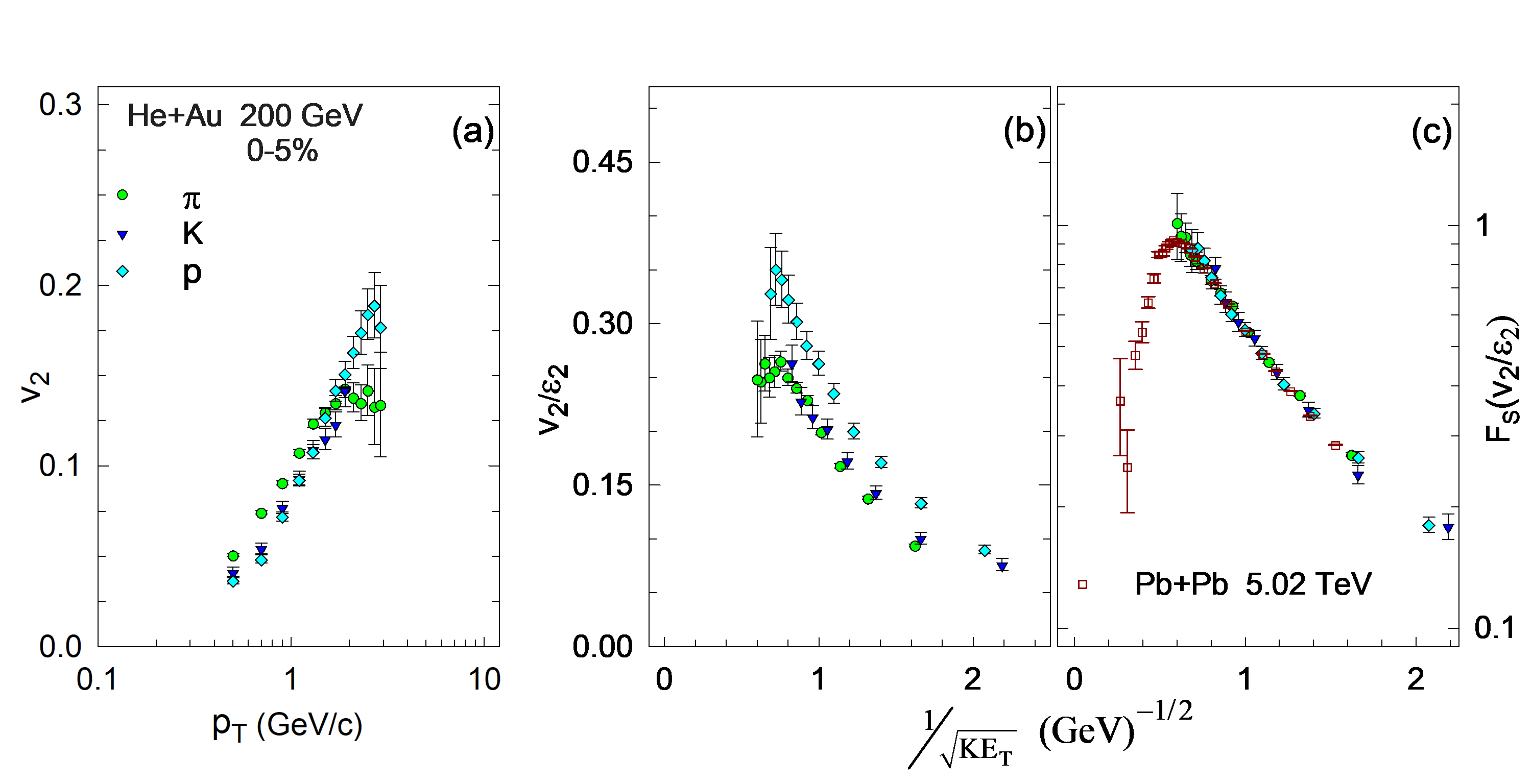}
	 \vskip -0.3 cm
  \caption{(Color online) Scaling procedure for 0–5\% central $^3$He+Au collisions at \(\sqrt{s_{NN}} = 200\)~GeV. 
Panel (a): raw \(v_2(p_T)\) for mesons (\(\pi^{\pm}, K^{\pm}\)) and baryons (\(p\bar{p}\)). 
Panel (b): eccentricity-scaled values \(v_2/\varepsilon_2\). 
Panel (c): final scaling function, compared with central Pb+Pb at \(\sqrt{s_{NN}} = 5.02\)~TeV. 
Data from ALICE and PHENIX~\cite{ALICE:2022zks,PHENIX:2017djs}.}
  \label{fig5}
\end{figure*}

Scaling functions were generated for Au+Au, Cu+Cu, and for small systems (p+Au, d+Au, $^3$He+Au) at \(\sqrt{s_{NN}} = 200\)~GeV. Figures~\ref{fig3} and \ref{fig5} show representative results for near-ultra-central (0--5\%) p+Au and $^3$He+Au. Panels (a) display raw \(v_2(p_T)\) for \(\pi^{\pm}\), \(K^{\pm}\), and \(p\bar{p}\). The expected species ordering is present, but splittings are smaller than in uc Pb+Pb and p+Pb (Figs.~\ref{fig1}, \ref{fig2}), reflecting lower RHIC energy densities, weaker pressure gradients, and diminished radial flow. Stronger hadronic re-scattering and enhanced viscous attenuation relative to Au+Au further limit species separation, indicating that hadronic transport plays a larger role in these systems. For Cu+Cu, peripheral centralities yield multiplicities comparable to $^3$He+Au, making it an important intermediate reference that bridges the transport behavior between asymmetric small systems and central Au+Au.

Panels (b) show \(v_2/\varepsilon_2\) vs.\ \(1/\sqrt{{\rm KE}_T}\), which accentuates baryon--meson separation in the flow-dominated region. Compared to Fig.~\ref{fig1}(b), the reduced proton blue shift signals weaker radial flow, more comparable to the small splitting in peripheral Pb+Pb (Fig.~\ref{fig2}b) and intermediate Cu+Cu than to the sizeable blue shifts seen in p+Pb (Figs.~\ref{fig1}e, \ref{fig2}e). The absence of full species convergence points to residual transport effects. Panels (c) present the fully scaled results: alignment improves at low \(p_T\), partly compensating for species differences, but deviations persist at intermediate and high \(p_T\), where limited opacity and short lifetimes reduce partonic energy loss and enhance non-flow. For comparison, central Pb+Pb at 5.02~TeV and Au+Au at 200~GeV are also shown, underscoring stronger collectivity and greater opacity in larger systems.

\begin{table}[h!]
\centering
\caption{Extracted scaling coefficients for Au+Au, Cu+Cu, p+Au, d+Au, and $^3$He+Au at \(\sqrt{s_{NN}} = 200\)~GeV. 
The value \(k_{\beta} = 0.63\) from Au+Au (0--10\%) serves as a baseline to compute the relative viscous attenuation 
\(k'_{\beta} = k_{\beta'}/k_{\beta}\) for the other systems. 
For Cu+Cu, a peripheral (40--50\%) selection yields multiplicities comparable to $^3$He+Au.}
\label{tab:coefficients2}
\begin{tabular}{lccccc}
\hline
System & Centrality (\%) & \(\gamma_{32}\) & \(k'_{\beta}\) & \(\zeta_{\rm hs}\) & \(\zeta_{\rm rf}\) \\
\hline
Au+Au (200 GeV)     & 0--10 & 0.05 & 1.00 & 0.08 & 0.29 \\
Cu+Cu (200 GeV)     & 40--50 & -0.44 & 1.12 & 0.14 & 0.03 \\
p+Au (200 GeV)      & 0--5 & -0.39 & 1.21 & 0.21 & 0.01 \\
d+Au (200 GeV)      & 0--5 & 0.00 & 1.30 & 0.14 & 0.06 \\
$^3$He+Au (200 GeV) & 0--5 & -0.29 & 1.25 & 0.14 & 0.05 \\
\hline
\end{tabular}
\end{table}

The deviations in Figs.~\ref{fig3}(c) and \ref{fig5}(c) follow a hierarchy p+Au \(>\) d+Au \(>\) $^3$He+Au, consistent with re-scattering dominance: smaller systems have larger surface-to-volume ratios and more localized hadronic interactions. Table~\ref{tab:coefficients2} confirms that collectivity in these systems is governed by hadronic re-scattering, not QGP dynamics. All three exhibit comparable viscous attenuation (\(k_{\beta}\)), minimal radial flow, and deviations consistent with suppressed jet quenching. Extending $^3$He+Au \(v_2(p_T)\) to higher \(p_T\) could clarify residual jet contributions. The intermediate Cu+Cu system, sampled at peripheral centrality, shows coefficients between Au+Au and the small asymmetric systems, reinforcing its role as a bridge in the system-size hierarchy. Comparisons with Au+Au and Pb+Pb highlight the transport evolution with increasing system size and energy density.

In central Au+Au at 200~GeV, low \(\eta/s\) (\(k_{\beta}\!\sim\!0.63\)), modest re-scattering (\(\zeta_{\rm hs}=0.08\)), and strong radial flow (\(\zeta_{\rm rf}=0.29\)) signal QGP-driven hydrodynamics with substantial jet quenching~\cite{PHENIX:2008saf}. Relative to Pb+Pb at 5.02~TeV, Au+Au shows reduced radial flow and lower effective \(\eta/s\), reflecting lower initial temperatures. A small but nonzero \(\gamma_{32}=0.05\) indicates only minor geometric deviations from the reference. Cu+Cu, sampled at peripheral centrality, exhibits coefficients intermediate in viscous attenuation but with re-scattering and flow values closer to the small asymmetric systems, underscoring its role as a bridge case.
 By contrast, RHIC small systems exhibit elevated \(\zeta_{\rm hs}\), suppressed \(\zeta_{\rm rf}\), and larger \(k'_{\beta}\), consistent with re-scattering–dominated collectivity. Their \(\gamma_{32}\) values, ranging from 0 to about –0.39, highlight system-dependent geometric effects tied to projectile structure and initial eccentricities. p+Au shows the strongest re-scattering and no radial flow; d+Au and $^3$He+Au develop small but finite flow from their larger size and geometry. With suppressed jet quenching and a lowered quenching threshold, these results confirm that RHIC small-system collectivity arises primarily from hadronic interactions, establishing a hadronic baseline for the onset of QGP-like behavior in larger or higher-energy systems.

In summary, this study analyzes azimuthal anisotropy scaling functions for identified mesons and baryons in large (Pb+Pb, Au+Au), intermediate (Cu+Cu), and small (p+Pb, p+Au, d+Au, $^3$He+Au) systems across a broad energy range. The extracted coefficients disentangle shear viscosity (\(\eta/s\)), radial flow, hadronic re-scattering, and jet quenching, revealing system-dependent dynamics. The fidelity of the scaling functions supports a hydrodynamic-like origin of anisotropy in the flow-dominated regime. Central Pb+Pb, Au+Au, and Cu+Cu exhibit strong radial flow, minimal re-scattering, and substantial jet quenching—signatures of long-lived QGP—whereas peripheral Pb+Pb and Cu+Cu reflect hadronic-dominated dynamics. In contrast, central p+Au, d+Au, and $^3$He+Au at RHIC show enhanced re-scattering, negligible flow, and absent quenching, consistent with hadronic collectivity. Ultra-central p+Pb at the LHC shows enhanced flow, reduced re-scattering, and small but nonzero quenching, indicative of partial QGP-like behavior. Together, these results demonstrate a transition from QGP- to hadronic-driven dynamics with decreasing system size and beam energy, and highlight the scaling framework as a powerful diagnostic of collectivity and medium evolution.

%
%%-----------------------------------------------------------
%\vspace{10pt}
%\section*{Acknowledgement}
%
%This research is supported by the US DOE under contract DE-FG02-87ER40331.A008. 
%%-----------------------------------------------------------
%\bibliographystyle{apsrev4-1}
%\bibliographystyle{elsarticle-num}
%\biboptions{sort&compress}
\bibliography{pid-refs-sys}

%%==============================================================================================
\end{document}